%
%
\magnification=1200
\tolerance=1400
\overfullrule=0pt
\baselineskip=11.5pt

\font\rmb=cmbx9 scaled \magstep 1
\font\rma=cmbx9 scaled \magstep 2
\font\rmm=cmbx9 scaled \magstep 3
 2
 2

\font\tenib=cmmib10
\font\sevenib=cmmib10 at 7pt
\font\fiveib=cmmib10 at 5pt
\newfam\mitbfam
\textfont\mitbfam=\tenib
\scriptfont\mitbfam=\sevenib
\scriptscriptfont\mitbfam=\fiveib

\mathchardef\bfo="0\the\mitbfam21
\mathchardef\om"0\the\bffam0A
\font\bigastfont=cmr10 scaled \magstep 3
\def\bdot{\hbox{\bigastfont .}}
\def\ueber#1#2{{\setbox0=\hbox{$#1$}%
  \setbox1=\hbox to\wd0{\hss$\scriptscriptstyle #2$\hss}%
  \offinterlineskip
  \vbox{\box1\kern0.4mm\box0}}{}}

\def\R{\rm I\kern-.18em R}
\def\E{\rm I\kern-.18em E}
\def\ref{\par\noindent\hangindent\parindent\hangafter1}

\def\1{_{\vert}}

\pageno=0
\topskip 1 true cm
\rightline{CERN--TH/99--165}
\bigskip\bigskip\bigskip
\centerline{\rmm On Average Properties of}
\bigskip
\centerline{\rmm Inhomogeneous Fluids in General Relativity}
\bigskip
\centerline{\rmm I: Dust Cosmologies}

\smallskip
\bigskip\bigskip
\centerline{\rma by}
\bigskip\bigskip
\centerline{\rma Thomas Buchert}
\vskip 1 true cm
\centerline{Theory Divison, CERN, CH--1211 Gen\`eve 23, Switzerland
\footnote*{on leave from:
Theoretische Physik, Ludwig--Maximilians--Universit\"at, 
Theresienstr. 37, D--80333 M\"unchen, Germany}}

\medskip

\centerline{e--mail: buchert@theorie.physik.uni-muenchen.de}


\vskip 2 true cm
\noindent
{\narrower
{\rma Summary:}
For general relativistic spacetimes filled with irrotational `dust' 
a generalized form of Friedmann's equations for an `effective'
expansion factor $a_{\cal D}(t)$ of inhomogeneous cosmologies is 
derived. Contrary to the standard Friedmann equations, which hold 
for homogeneous--isotropic cosmologies, the new equations include 
the `backreaction effect' of inhomogeneities on the average expansion 
of the model. A universal relation between `backreaction' and average 
scalar curvature is also given. For cosmologies whose averaged spatial 
scalar curvature is proportional to $a_{\cal D}^{-2}$, the expansion 
law governing a generic domain can be found. However, as the general 
equations show, `backreaction' acts as to produce average curvature 
in the course of structure formation, even when starting with space 
sections that are spatially flat on average. 

}

\vfill\eject

\topskip 0 true cm

\noindent{\rmm 1. The Averaging Problem}
\bigskip\bigskip
\noindent
A longstanding question in cosmology is, how to average a general
inhomogeneous model (Ellis 1984). 
Also one would like to know under which assumptions, if any, the average
variables obey Friedmann's equations lying at the basis of any
theory for structure formation in the Universe.
An answer for cosmologies containing a pressure--free fluid (`dust')
has been given recently in the framework of 
Newtonian cosmology (Buchert \& Ehlers 1997;
see the references therein for the formulation of the averaging problem and earlier attempts 
to solve it). Some results relevant to the present work may be briefly summarized as 
follows: Consider any simply--connected, compact spatial domain in
Euclidean space ${\cal D} \subset \E^3$ with volume $|{\cal D}| = V_{\cal D}$. 
Then, the spatial average of Raychaudhuri's equation for the
evolution of the expansion rate, 
under the assumption of conservation of the
domain's mass, yields an equation for the scale factor 
$a_{\cal D}\propto V_{\cal D}^{1/3}$ (which depends on 
content, shape and position of the spatial domain). This equation contains as source terms, 
besides the average mass density, averages over fluctuations of the shear, vorticity 
and expansion scalars due to the presence of
inhomogeneities. These `backreaction' terms vanish, 
if the average is performed over the whole universe having
topologically closed space sections. For vanishing `backreaction' this equation
is equivalent to the 
standard Friedmann equation for homogeneous--isotropic universes.
\smallskip
In the present paper I provide the corresponding answer in the 
framework of general 
relativity. A geometrical relation, having no Newtonian analogue, 
delivers additional information that allows to obtain a single equation 
relating the `backreaction' and average curvature terms. This equation
holds for any spatial domain and for a large class of inhomogeneous cosmologies 
with curved space sections without perturbative assumptions.
The solution of the `backreaction problem' for scalar characteristics 
can be found in the case of
space sections whose Ricci scalar averages out to zero, or, 
displays a dependence $\propto a_{\cal D}^{-2}$ as in the standard
model, which implies that the averaged scalar curvature 
decouples from the `backreaction' term. However, in general,
`backreaction' due to the presence of inhomogeneities impacts on the
averaged Ricci scalar in the course of structure formation.
A general solution to this problem would incorporate a scale--dependent 
description of the density, the expansion and other scalar
variables of any structure formation model. 
\smallskip
In Section 2 we proceed by prescribing the basic equations and the averaging 
procedure;  
then we give the general equations governing the domain--dependent 
scale factor $a_{\cal D}(t)$ in a {\it Theorem}.
An equivalent set of these equations presented in {\it Corollary 1} demonstrates
the surprising result that, inspite of the non--commutativity of averaging and
time evolution, the averaged quantities obey the same equations as the local
ones. Inferred from the averaged equations {\it Corollary 2} defines 
a set of average characteristics in analogy to the cosmological parameters
of the standard model. A compact form of the previous results is presented in
{\it Corollary 3}, which displays a universal relation between average scalar 
curvature and `backreaction'.  
Finally, in Section 3, we discuss
some immediate consequences of this result, and comment on related work. 
In Appendix A we present the Newtonian analogues, while Appendix B
gives some illustrative examples of expansion laws. Appendix C is dedicated
to an alternative derivation of the averaged equations. 

\bigskip\bigskip\bigskip

\noindent{\rmm 2. Averaging Einstein's Equations for Scalars}
\smallskip
\bigskip\bigskip
\noindent
{\rma 2.1. Choice of foliation and basic equations}
\bigskip\medskip\noindent
We shall restrict ourselves to the case of irrotational fluid motion
with the simplest matter model `dust' (i.e. vanishing pressure).
In this case the flow is geodesic and space--like hypersurfaces can be
constructed that are flow--orthogonal at every spacetime event in a $3+1$ 
representation of Einstein's equations.
\smallskip
We start with Einstein's equations$^1$\footnote{}{$^1$Greek indices run through
$0 ... 3$, while latin indices run through $1 ... 3$; summation over 
repeated indices is understood. A semicolon will denote covariant derivative with 
respect to the 4--metric with signature $(-,+,+,+)$; the units are such that $c=1$.}
$$
R_{\mu\nu} - {1\over 2}g_{\mu\nu}R = 8\pi G \varrho u_{\mu} u_{\nu}
-\Lambda g_{\mu\nu} \;\;\;, 
\eqno(1a)
$$
with the Ricci tensor $R_{\mu\nu}$, its trace $R$, the fluid's $4-$velocity 
$u^{\mu}$ ($u^{\mu}u_{\mu} = -1$), the cosmological constant $\Lambda$,
and the rest mass density $\varrho$ obeying the conservation law
$$
\left( \varrho u^{\mu} u^{\nu}\right)_{\;;\mu} \;=\;0\;\;\;.\eqno(1b)
$$ 
We choose a flow--orthogonal coordinate system
$x^{\mu} = (t,X^k)$ (i.e., Gaussian or normal coordinates which are comoving
with the fluid). Writing $x^{\mu}=f^{\mu}(X^k ,t)$ we have 
$u^{\mu}={\dot f}^{\mu} = (1,0,0,0)$ and 
$u_{\mu}={\dot f}_{\mu} = (-1,0,0,0)$,
where the dot denotes partial derivative with respect to proper time $t$.
 
These coordinates are defined such as to label
geodesics in spacetime, i.e., $u^{\nu}u^{\mu}_{\;\,;\nu}=0$. With the 
choice of vanishing $3-$velocity the coordinates are in addition 
chosen to be {\it comoving}.
Thus, in a $3+1$--splitting of spacetime, the spatial set of Gaussian 
coordinates also 
labels fluid elements or trajectories in 3--space, ${\dot X}^k = 0$; we are
entitled to call $X^k$ {\it Lagrangian coordinates}, because
they are identical to those in classical fluid dynamics. 
It should be emphasized, however, that the final result will be covariant with 
respect to the given foliation of spacetime and thus not dependent on this particular
choice of coordinates. 
\medskip\medskip
Let $(t,X^k)$ be the independent variables. As dependent variables we may
choose the spatial $3-$metric
${g}_{ij}$ (the first fundamental form of the hypersurfaces of 
constant $t$) in the line element
$$
ds^2 = - dt^2 + {g}_{ij} dX^i dX^j \;\;,\eqno(1c)
$$
the extrinsic curvature tensor $K_{ij}:=-h^{\alpha}_{\;\,i}h^{\beta}_{\;\,j} 
u_{\alpha;\beta}$ 
(the second fundamental form
of the hypersurfaces of constant $t$) with the projection tensor into the
hypersurfaces orthogonal to $u_{\alpha}$,  
$h_{\alpha\beta}:= g_{\alpha\beta} + u_{\alpha}u_{\beta}$, and the rest mass density $\varrho$.
Einstein's equations (1a) together with the continuity equation (1b) 
(contracted with $u_{\nu}$) then are equivalent to the following system 
of equations (see, e.g., Arnowitt et al. 1962, York 1979)
consisting of the {\it constraint equations}$^2$\footnote{}
{$^2$In Eq. (2b) $_{||}$ denotes covariant 
derivative with respect to the $3-$metric
${g}_{ij}$, while a single vertical slash denotes partial derivative 
with respect to the Lagrangian coordinates $X^i$; note that in the present
case the covariant spatial derivative of a scalar 
reduces to the partial derivative.
The overdot denotes partial time--derivative (at constant $X^i$) as before.}
$$\eqalignno{
{1\over 2}\left( {\cal R} 
+ K^2 - K^i_{\,\;j} K^j_{\,\;i} \right) 
&= 8\pi G\varrho + \Lambda \;\;,&(2a)\cr
K^i_{\,\;j || i} - K_{| j} &= 0 \;\;,&(2b)\cr}
$$
and the {\it evolution equations} for the density and the two fundamental
forms:
$$\eqalignno{
{\dot \varrho} &=  K \varrho \;\;,&(2c)\cr
\left({g}_{ij}\right)^{\bdot} &= -2 \;\,{g}_{ik}K^k_{\,\;j} \;\;,&(2d)\cr
\left({K}^i_{\,\;j}\right)^{\bdot} &=  K K^i_{\,\;j} + {\cal R}^i_{\,\;j} - 
(4\pi G \varrho + \Lambda)\delta^i_{\,\;j} \;\;.&(2e)\cr}
$$ 
${\cal R}:= {\cal R}^i_{\,\;i}$ and $K:=K^i_{\,\;i}$ 
denote the traces of the spatial Ricci tensor ${\cal R}_{ij}$
and the extrinsic curvature tensor $K_{ij}$, respectively. 
Expressing the latter in terms of kinematical quantities,
$$
-K_{ij} = \Theta_{ij} = \sigma_{ij} + {1\over 3}\theta g_{ij}\;\;\;;\;\;\;-K = \theta 
\;\;\;,\eqno(3)
$$
with the expansion tensor $\Theta_{ij}$, the trace--free symmetric shear tensor $\sigma_{ij}$ 
and the expansion rate $\theta$, we may write Eqs. (2) in the form
$$\eqalignno{
{1\over 2} {\cal R}
+ {1\over 3}\theta^2 - \sigma^2  
&= 8\pi G\varrho + \Lambda \;\;,&(4a)\cr
{\sigma}^i_{\,\;j || i} &= {2\over 3} \theta_{| j} \;\;,&(4b)\cr
{\dot \varrho} &= - \theta \varrho \;\;,&(4c)\cr
\left({g}_{ij}\right)^{\bdot} &= 2 \;\,{g}_{ik}{\sigma}^k_{\,\;j} 
+ {2\over 3}\theta {g}_{ik}{\delta}^k_{\,\;j}\;\;,&(4d)\cr
\left({\sigma}^i_{\,\;j}\right)^{\bdot} &= - \theta {\sigma}^i_{\,\;j} - 
{\cal R}^i_{\,\;j} +{2\over 3}{\delta}^i_{\,\;j}\left[\sigma^2 - {1\over 3}
\theta^2 + 8\pi G\varrho + \Lambda \right] \;\;,&(4e)\cr}
$$
where we have introduced the rate of shear $\sigma^2 := {1\over 2}\sigma^i_{\;\,j} 
\sigma^j_{\;\,i}$.
To derive this last equation we have used Raychaudhuri's equation
$$
\dot\theta + {1\over 3}\theta^2 + 2\sigma^2 + 4\pi G\varrho - 
\Lambda \;=\;0 \;\;,\eqno(5a)
$$
which follows from the trace of Eq. (4e) combined with the constraint (4a).
Using again the constraint (4a), we may cast the trace--free part 
(4e) into the form
$$
\left({\sigma}^i_{\,\;j}\right)^{\bdot} + \theta {\sigma}^i_{\,\;j} = -
\left({\cal R}^i_{\,\;j} -{1\over 3}{\delta}^i_{\,\;j}{\cal R}\right)
\;\;.\eqno(5b)
$$

\medskip\noindent 
This set of equations has recently been discussed in connection with
perturbation theory by Kasai (1995), Matarrese (1996, and ref. therein) 
and by Matarrese \& Terranova (1996), as well as in the papers by
Russ et al. (1996, 1997).
Here, we proceed without perturbation theory.
\bigskip\noindent
Taking the trace of Eq. (2d), written in the form 
$$
K^i_{\,\;j} = -{1\over 2}{g}^{ik}\left({g}_{kj}\right)^{\bdot}\;\;\;,
$$
and defining 
$$
J(t, X^i):= \sqrt{\det({g}_{ij})} \;\;\;,\eqno(6a)
$$
we obtain with ${1\over 2}{g}^{ik}\left({g}_{ki}\right)^{\bdot}
= (\ln J)^{\bdot}$ the identity
$$
\dot J = - K J = \theta J\;\;\;.\eqno(6b)
$$
Using it we can integrate the continuity equation for the rest mass density
(2c) along the flow lines:
$$
\varrho (t, X^i) = (\varrho (t_0 , X^i) J (t_0 , X^i )) J^{-1} \;\;\;.\eqno(7)
$$
\bigskip
\noindent 
Both, Raychaudhuri's equation (5a) and the integral of the continuity 
equation (7) are identical to their Newtonian counterparts
(Buchert \& Ehlers 1997).
Below we shall make explicit use of the constraint (4a), which has no 
Newtonian analogue. This equation 
will provide a key element for the understanding of the `backreaction' problem. 

\bigskip\noindent
For later discussion it is convenient to also introduce the abbreviations 
$\bf I$ and $\bf II$ 
for two of the scalar invariants of the expansion tensor,
its trace, 
$$
{\bf I}: = \Theta^{\ell}_{\,\;\ell} = \theta \;\;\;,\eqno(8a)
$$
and the dispersion of its diagonal components,  
$$
{\bf II}: = {1\over 2}\left(\theta^2 - \Theta^{\ell}_{\,\;k}\Theta^{k}_{\,\;\ell}\right)
= {1\over 3}\theta^2 - \sigma^2
\;\;\;.\eqno(8b)
$$

\vfill\eject

\noindent
{\rma 2.2. Averaging the traces of Einstein's equations}
\bigskip\medskip\noindent
Spatially averaging equations for scalar fields 
is a covariant operation given a foliation of spacetime.
Therefore, as we already pointed out in the paper on averaging the 
Newtonian equations
(Buchert \& Ehlers 1997), we may average, e.g., Raychaudhuri's equation (5a)
in full formal analogy to the Newtonian case.

Let us define the averaging operation. Spatial averaging of a scalar field
$\Psi$ as a function of Lagrangian coordinates and time 
on an arbitrary compact portion of the fluid $\cal D$ is straighforward$^3$ 
\footnote{}{$^3$This averaging method functionally depends on content, shape and
position of the spatial domain of averaging, which we consider as being 
given (see: Stoeger et al. 1999 for alternative averagers).
It should be stressed that we do not attempt to average the spacetime
geometry as a whole; useful information for cosmology may be already 
obtained by averaging the scalar parts.} 
and is defined by the volume integral
$$
\langle \Psi (t, X^i)\rangle_{\cal D}: = 
{1\over V_{\cal D}}\int_{\cal D} J d^3 X \;\;\Psi (t, X^i) \;\;\;,\eqno(9a)
$$
with the volume element $dV:= J d^3 X$ of the spatial hypersurfaces of constant 
time. The volume itself is given by$^4$\footnote{}{$^4$In comparison with the 
Newtonian definition $V_{\cal D} = \int_{{\cal D}(t)} d^3 x 
= \int_{{\cal D}(t_0)} J d^3 X$, where $x_i$ are Eulerian and $X_i$ Lagrangian
coordinates, our domain $\cal D$ corresponds to a Lagrangian domain, because it is
transported along geodesics; however, in contrast to the Newtonian case,
it is implicitly time--dependent due to the evolution of the 3--metric.}
$$
V_{\cal D}(t) : = \int_{\cal D} J d^3 X \;\;\;.\eqno(9b)
$$
We also introduce a dimensionless (`effective') 
scale factor via the volume (normalized by the volume of the initial
domain $V_{{\cal D}_o}$), 
$$
a_{\cal D}(t) : = \left({V_{\cal D}(t)\over V_{{\cal D}_o}}\right)^{1/3} \;\;\;.\eqno(9c)
$$
\smallskip
\noindent
Thus, the averaged expansion rate may be written in terms of the scale factor:
$$
\langle\theta\rangle_{\cal D} = {{\dot V}_{\cal D}\over V_{\cal D}} = 3 
{{\dot a}_{\cal D} \over a_{\cal D}} \;\;\;.\eqno(9d)
$$
The integral (7) states the conservation of the total rest mass $M_{\cal D}$ within a portion 
of the fluid $\cal D$
as it is transported along the flow lines,
$$
M_{\cal D} = \int_{\cal D} J d^3 X\;\varrho \;\;=\;\; const.\;\;
\Leftrightarrow \;\;\langle\varrho\rangle_{\cal D} = 
{M_{\cal D} \over V_{{\cal D}_o}a_{\cal D}^3 }\;\;\;.\eqno(9e)
$$ 
Employing this averaging procedure we may easily prove many 
statements found in (Buchert \& Ehlers 1997) which also hold in general 
relativity. From these results we are going to use the `Commutation rule' (here
written for an arbitrary scalar field $\Psi$):

\bigskip\bigskip\noindent
\underbar{\bf Lemma$\;\;$({\it Commutation rule})} 
\bigskip\noindent
$$
\langle \Psi\rangle_{\cal D}^{\bdot} - \langle{\dot \Psi}\rangle_{\cal D}
= \langle \Psi\theta\rangle_{\cal D} - 
\langle \Psi\rangle_{\cal D}\langle\theta\rangle_{\cal D}\;\;\;.\eqno(9f)
$$
\smallskip\bigskip\noindent
Averaging the Hamiltonian constraint (4a) and Raychaudhuri's equation (5a) 
with the help of the prescribed procedure, 
we end up with the following two 
equations, which we may summarize in the form of a theorem.

\bigskip\bigskip
\noindent
\underbar{\bf Theorem $\;\;$({\it Equations for the effective scale factor})}
\bigskip\noindent
The spatially averaged equations for the scale factor $a_{\cal D}$, 
respecting mass conservation, read:
\medskip\noindent
averaged Raychaudhuri equation:
$$
3{{\ddot a}_{\cal D} \over a_{\cal D}} + 4\pi G 
{M_{\cal D}\over V_{{\cal D}_o}a_{\cal D}^3} - \Lambda \;=\; {\cal Q}_{\cal D}
\;\;\;;\eqno(10a)
$$
averaged Hamiltonian constraint:
$$
3\left( {{\dot a}_{\cal D}\over a_{\cal D}}\right)^2 - 8\pi G
{M_{\cal D}\over V_{{\cal D}_o}a_{\cal D}^3} + {1\over 2}\langle {\cal R} \rangle_{\cal D} 
- \Lambda = -{{\cal Q}_{\cal D}\over 2} 
\;\;\;,\eqno(10b)
$$
where the mass $M_{\cal D}$, the averaged spatial Ricci scalar 
$\langle {\cal R}\rangle_{\cal D}$ and 
the `backreaction' ${\cal Q}_{\cal D}$ are domain--dependent spatial constants 
and, except the mass, time--dependent functions. In particular, 
the `backreaction' source term is given by 
$$
{\cal Q}_{\cal D} : = 2 \langle{\bf II}\rangle_{\cal D} 
- {2\over 3}\langle {\bf I}\rangle_{\cal D}^2 \;=\;
{2\over 3}\langle\left(\theta - \langle\theta\rangle_{\cal D}\right)^2 
\rangle_{\cal D} - 2\langle\sigma^2\rangle_{\cal D}
\;\;\;.\eqno(10c)  
$$

\bigskip\bigskip\bigskip
\noindent
We also note the following surprising property of the averaged equations 
compared with their local forms: in spite of the non--commutativity of 
the averaging procedure and the dynamical evolution, which is expressed by the
Commutation rule (9f), we find that the same equations hold for the 
averaged and the local quantities provided we express them in terms of
the invariants (8). This establishes the following

\bigskip\bigskip
\noindent
\underbar{\bf Corollary 1 $\;\;$({\it Averaged equations})}
\bigskip\noindent
The spatial averages of the Hamiltonian constraint (4a), the continuity equation
(4c) and Raychaudhuri's equation (5a) read:
$$
\eqalignno{
{1\over 2} \langle{\cal R}\rangle_{\cal D} 
&= 8\pi G\langle\varrho\rangle_{\cal D} + \Lambda - 
\langle{\bf II}\rangle_{\cal D} \;\;,&(11a)\cr
\langle\varrho\rangle^{\bdot}_{\cal D} &= - \langle\theta\rangle_{\cal D} 
\langle\varrho\rangle_{\cal D}\;\;,&(11b)\cr
\langle\theta\rangle_{\cal D}^{\bdot}  
&=-\langle\theta\rangle_{\cal D}^2 + \Lambda - 
4\pi G\langle\varrho\rangle_{\cal D} + 2 \langle{\bf II}\rangle_{\cal D}
\;\;\;,&(11c)\cr}
$$
i.e., the averages $\langle\varrho\rangle_{\cal D}$, 
$\langle\theta\rangle_{\cal D}$,
$\langle{\cal R}\rangle_{\cal D}$ and $\langle{\bf II}\rangle_{\cal D}$
obey the same equations as the local fields $\varrho$, $\theta$, 
$\cal R$ and $\bf II$.
(The reason for this nontrivial property is the special type of nonlinearities
featured by the gravitational system, e.g. the nonlinearity 
in $\theta$ contained in Raychaudhuri's equation.)

\vfill\eject
\noindent
\underbar{\bf Corollary 2 $\;\;$({\it Dimensionless characteristics of 
inhomogeneous cosmologies})}
\bigskip\noindent
As in the standard homogeneous--isotropic cosmologies we may 
introduce a domain--dependent
Hubble function $H_{\cal D}: ={{\dot a}_{\cal D}\over a_{\cal D}}$, and 
dimensionless average characteristics 
as follows:
$$
\eqalignno{
\Omega_m : &= {8\pi G M_{\cal D} \over 3 V_{{\cal D}_o}a_{\cal D}^3 H_{\cal D}^2 }  \;\;,&(12a)\cr
\Omega_{\Lambda} :&= {\Lambda \over 3 H_{\cal D}^2 }\;\;,&(12b)\cr
\Omega_{k} :&= - {\langle {\cal R}\rangle_{\cal D}\over 6 H_{\cal D}^2 }\;\;,&(12c)\cr
\Omega_{\cal Q} :&= - {{\cal Q}_{\cal D} \over 6 H_{\cal D}^2 }
\;\;\;,&(12d)\cr}
$$
which, in view of (10b), obey
$$
\Omega_m\;+\;\Omega_{\Lambda}\;+\;\Omega_{k}\;+\; \Omega_{\cal Q}\;=1\;\;.\eqno(12e)
$$
All these dimensionless ``cosmological parameters'' actually depend 
on the spatial scale of averaging including the 
dimensionless cosmological constant, which depends on scale through $H_{\cal D}$.  
\bigskip
The equations (10a,b) form a system of two equations for the three
unknown variables $a_{\cal D}$, $\langle{\cal R}\rangle_{\cal D}$ and
${\cal Q}_{\cal D}$. Therefore, we cannot solve the `backreaction' problem
for scalars based on this system. 
We may eliminate the
`backreaction' term from (10a) and insert it into (10b). 
This results in an equation for the scale factor
$a_{\cal D}$ and the average Ricci scalar of the domain.
Alternatively we may proceed as follows: we calculate the time--derivative 
of Eq. (10b) and insert
into the resulting equation (10a) and (10b). 
This yields a universal relation between the averaged Ricci scalar 
and the `backreaction' term: 
\bigskip\bigskip
\noindent
\underbar{\bf Corollary 3 $\;\;$({\it Relation between average scalar 
curvature and `backreaction'})}
\bigskip\noindent
A necessary condition of integrability of Eq. (10a) to yield 
Eq. (10b) is provided by the relation:
$$
{\cal Q}^{\bdot}_{\cal D} + 6 {{\dot a}_{\cal D}\over a_{\cal D}} 
{\cal Q}_{\cal D} +  
\langle {\cal R} \rangle_{\cal D}^{\bdot}
+ 2 {{\dot a}_{\cal D}\over a_{\cal D}} \langle {\cal R} \rangle_{\cal D}\;=\;0
\;\;\;,\eqno(13a)
$$ 
or, equivalently,
$$
\left(a_{\cal D}^6 {\cal Q}_{\cal D}\right)^{\bdot} + 
a_{\cal D}^4 \left(a_{\cal D}^2 \langle {\cal R} 
\rangle_{\cal D}\right)^{\bdot} \;=\;0\;\;.\eqno(13b)
$$
\smallskip\noindent
In Appendix B we give some examples of expansion laws that can be derived
from this relation.

\vfill\eject
\noindent
\underbar{\bf Notes:}
\bigskip\noindent
Eq. (10a) has been already confirmed as a valid equation in general relativity in 
the work on averaging the Newtonian equations (Buchert \& Ehlers 1997);
an equation analogous to Eq. (10b) has been 
derived by Carfora \& Piotrkowska (1995) in connection with manifold deformations at one instant 
using the constraint equations of general relativity.
In our derivation of Eq. (10b) we have inserted the `backreaction' term ${\cal Q}_{\cal D}$ 
back into the constraint equation (4a) and have used (9d).

\smallskip
Russ et al. (1997) have also used the equations (10a,b)  
(in a truncated form and using a reference background solution) for 
the purpose of perturbative calculations of the `backreaction' effect.  
Note that, contrary to their derivation, 
we have not performed a conformal rescaling of the
metric, nor have we used the splitting into a background reference solution
and deviations thereof. 
We postpone further comments on their work to Section 3.

\bigskip\bigskip\bigskip

\noindent{\rmm 3. Discussion and Perspectives}
\bigskip\bigskip
\noindent
{\rma 3.1. Summary of results}
\bigskip\noindent
We have derived a generalized form of Friedmann's differential equations 
including `backreaction terms' due to the presence of inhomogeneities.
One of these equations was obtained on the basis of averaging 
Raychaudhuri's equation
on spatial domains whose mass content is preserved in time.
It is formally identical to the equation derived in the framework of Newtonian 
cosmology (Appendix A). The other equation arises by averaging the Hamiltonian 
constraint (having no Newtonian analogue). It delivers  
an additional relation between the averaged spatial Ricci curvature scalar and 
the `backreaction term'. We have shown that there exist special 
solutions which describe the evolution of the average curvature and 
`backreaction' terms (Appendix B). 
\smallskip\noindent
Let us list some immediate conclusions which may be drawn:

\smallskip\noindent$1.$ 
The average expansion of inhomogeneous cosmologies does,
in general, not follow the expansion law of the standard FRW cosmologies.
There are, however, generalized expansion laws which govern the
motion of arbitrary spatial domains, provided assumptions on the relation between
the averaged Ricci scalar and the `backreaction term' are made. Here, perturbation theory
would be useful to establish such relations. 

\smallskip\noindent$2.$ 
The general expansion law shows that `backreaction'
due to the presence of inhomogeneities implies the existence of
a non--vanishing average Ricci scalar in general situations. This is true 
even if we consider domains which are on average Ricci flat at some initial
instant.

\medskip
Some comments about these conclusions are in order.

\medskip\noindent$ad 1.$ We may stipulate that the assumption of 
vanishing average Ricci scalar could be a sensible one, if we 
consider {\it typical} portions of the Universe (which itself may have on average 
flat space sections to a good approximation). As demonstrated in Appendix B a solution 
can be obtained in this case and the evolution of the average expansion is
then exactly known. Consideration of the general case reveals, however, that 
this point of view is too naive: Looking at Eq.~(10b) we must expect
that the dimensionless contribution to the averaged scalar curvature
$\langle {\cal R}\rangle_{\cal D}/G\langle\varrho\rangle_{\cal D}$ might be 
of the same order as the dimensionless contribution to the `backreaction'
${\cal Q}_{\cal D}/G\langle\varrho\rangle_{\cal D}$. Approximating 
the average curvature by zero relies on a similar prejudice as saying that 
`backreaction' may be neglected. {\it Both approximations imply restrictions on
general inhomogeneous models.}

\medskip\noindent$ad 2.$ Here, we may imagine the likely situation that 
an initially critical universe in the sense of the 
Einstein--de Sitter model may evolve into an under-- or overcritical universe,
respectively, in the
course of structure formation. 
Hence, it is possible that a Ricci flat universe  
develops into a universe with on average negative/positive spatial curvature
at the present epoch. 
From the point of view of the standard inflationary paradigm
the former situation is favoured, when the theoretical expectation of an on average
Ricci flat universe at the exit epoch is compared with 
measurements of the density parameter at the present epoch.

\medskip\medskip
For the general case a solution seems to lie beyond the scope of this article.
Let us illustrate why the system of averaged equations (13a,b) cannot be closed
on the level of scalars and also, how we may achieve closure by additional
assumptions.

\bigskip\bigskip
\noindent{\rma 3.2. Attempting closure of the averaged equations}
\bigskip\noindent
In order to obtain a more general result, we would like to find 
an independent evolution equation for the spatial Ricci scalar. 
Kasai (1995, appendix) has derived an evolution equation for the spatial 
Ricci tensor. His equation reads$^5$\footnote{}{$^5$Note that we have used
the canonical definition of the extrinsic curvature in this paper: 
$K_{ij} = -\Theta_{ij}$.}:
$$
\left({\cal R}^i_{\;j}\right)^{\bdot} - 2 K^i_{\;\ell} {\cal R}^{\ell}_{\;j} 
\,=\,
-K^{i\;\;\;\; ||\ell}_{\;\ell || j} - K^{\ell \;\; || i}_{\;j\;\;\; || \ell} + 
K^{i\;\; ||\ell}_{\;j\;\;\; ||\ell}
+ K^{\ell\;\;||i}_{\;\ell\;\;\; ||j}\;\;\;.\eqno(14a)
$$
This relation is purely geometrical and makes no use of the field equations.
Taking the trace of Eq. (14a) we first obtain
$$
{1\over 2}{\dot{\cal R}} - K^{\sigma}_{\;\;\ell}{\cal R}^{\ell}_{\;\;\sigma} 
= -\left( K^{\ell}_{\;\;\sigma || \ell} - K_{| \sigma}\right)^{|| \sigma} 
\;\;\;.\eqno(14b)
$$
The r.--h.--s. of this equation vanishes according to the momentum constraints 
(2b).
Using these constraints, the field equation (2e) to eliminate 
${\cal R}_{ij}$ in favour of $K_{ij}$ and the definition (8b) we obtain with 
$K_{ij} = - \Theta_{ij}$:
$$
{1\over 2} {\dot{\cal R}} = {\dot\theta}\theta + \theta^3 - {\dot{\bf II}}
- 2\theta{\bf II} - \theta(4\pi G\varrho + \Lambda) \;\;.\eqno(14c)
$$
Inserting Raychaudhuri's equation (5a) we find that Eq. (14c) 
is just the time--derivative of the Hamiltonian constraint (4a)
combined with (4c). Hence, the trace of the evolution equation (14a) 
for the spatial  
Ricci tensor cannot be used to close the system of equations
and to solve the `backreaction' problem.

We may try to use Eq. (5b), which is an equation for the trace--free parts,
and contract this equation into a relation among scalar quantities.
Indeed, if we contract Eq. (5b) with $\sigma^j_{\;\,i}$ and eliminate the
expression ${\cal R}^i_{\;\,j}\sigma^j_{\;\,i}$ from Eq. (14b) we obtain
$$
{1\over 2}\dot{\cal R} + {1\over 3}\theta{\cal R} 
= (\sigma^2)^{\bdot} + 2\theta \sigma^2 \;\;.\eqno(14d)
$$
Using the Hamiltonian constraint (4a), Eq. (14d) can be written as
$$
\dot{\cal R} + \theta {\cal R} - 2\theta \Lambda + 2 \dot{\bf II} + 
2\theta {\bf II} \;=\;0\;\;,\eqno(14e)
$$
which, however, can also be obtained by inserting the Hamiltonian constraint into 
its time--derivative.

These examples show that the use of any scalar part of Einstein's equations
will not give a closed system of {\it ordinary} differential equations
(see also Kofman \& Pogosyan 1995).
For the averaged variables we can also not expect this in view of
{\it Corollary 1}:
it states the equivalence of the
averaged dynamics to the dynamics for the local field quantities.
If this could be 
achieved in full generality, then this would be equivalent to having solved the 
full Einstein dynamics for the scalar parts, since the size of the domains 
could be arbitrarily chosen. 
 
\bigskip\bigskip
\noindent{\rma 3.3. Expansion law for closed universe models ?}
\bigskip\noindent
As discussed above an effort beyond the scope of ordinary differential 
equations for scalars is needed to close the system of 
equations for the average dynamics on any spatial domain. 

Notwithstanding, such an effort can be successful, if further constraints on the average dynamics
are considered, most notably integral constraints which restrict the 
morphological characteristics of domains. Among them the integrated 
curvatures and, in particular, topological constraints that arise by 
restricting the Euler--characteristic of the surfaces bounding the domains.
We already implied the topological constraint that the domains over which we average have to be
simply--connected. In Newtonian cosmology (Buchert \& Ehlers 1997) 
we have established a global criterion: if we extend
this simply--connected domain to the whole Universe having topologically closed space sections
(e.g., toroidal models), then this results in ${\cal Q}_{\cal D} = 0$ on the closure scale.
It is therefore to be expected that such a constraint may also close the system of averaged 
equations in 
general relativity. We do not necessarily have the simple Newtonian condition. 
{\it Corollary 3} states
a general connection between ${\cal Q}_{\cal D}$ and $\langle{\cal R}\rangle_{\cal D}$: 
a vanishing `backreaction' would imply that all the contributions 
of the local curvatures that are produced
by the inhomogeneities obey the ``conspiracy'' to sum up
to the standard value $\propto a(t)^{-2}$
(where $a(t)$ is a solution of a standard FRW cosmology).  

In curved spacetimes it is not straightforward to establish such a constraint and
the line of arguments given in the
Newtonian treatment is not conclusive in the present context. 
To illustrate this statement let us
look at the extrinsic curvature tensor according to its definition as the ($4-$dimensional)
covariant spatial derivative of the 4--velocity. 
Invariants built from $K_{ij}$ and, consequently,
the expression ${\cal Q}_{\cal D}$, cannot be written as total covariant divergences 
of vector fields {\it in} the
hypersurfaces. As an example we look at the trace of $K_{ij}\;$, 
$K=-u^{\alpha}_{\;\,;\alpha} =-\theta$;
the value of $K$ on the hypersurfaces is covariantly defined, but the vector field 
$u^i$ vanishes according to our spacetime foliation. 
A similar problem arises in the case of the second invariant.
 
\smallskip
Forthcoming efforts should be directed towards finding 
a topological closure condition for the hypersurfaces in order to determine
the global average properties of the world models. 
This problem is more involved, since we 
cannot expect that the domains remain simply--connected.
Working in a 4--dimensional tube of spacetime
that is bounded by space--like hypersurfaces and considering the limit
of vanishing distance between these hypersurfaces, Yodzis (1974) attempted to
derive average properties of {\it closed} space sections. His argument is reviewed
in Appendix C, where it is shown that topological restrictions do not enter
and his result holds for arbitrary compact and simply--connected domains.

\bigskip 
We conclude:

\smallskip\noindent$3.$ 
We were not able to produce an argument 
analoguous to the Newtonian treatment stating that the `backreaction term'  
vanishes for topologically closed space 
sections, if integrated over the whole space. 
Without such an argument averaged inhomogeneous cosmologies cannot be identified
with the standard FRW cosmologies on any spatial scale.
To justify this identification as an approximation there is presently 
no sufficiently general quantitative result as to whether the `backreaction' 
term could be 
neglected on some scale or, in words suggested by {\it Corollary 3}, whether 
the averaged curvature decouples from the inhomogeneities.

\bigskip\bigskip
\noindent
{\rma 3.4. Remarks on perturbation theory}
\bigskip\noindent
I stated above that until present we don't know any quantitative calculation
which may justify neglection of the `backreaction' term on some scale.
The reader may object that there exist several approximate calculations of
the `backreaction' effect in perturbation theory. 
However, there are severe obstacles for perturbative calculations which we 
are going to discuss now.
\smallskip
As an example I would like to 
comment on a recent calculation by Russ et al. (1997; see also the references
therein): based on the system of equations (10a), (10b) (using a rescaling 
of the metric and a split into background and perturbations)
the `backreaction' term was calculated within a second--order perturbation
approach. In order to make the calculations concrete, Russ et al.
have assumed periodic boundary conditions on some (very large) domain. 
Looking at their expression for the `backreaction' term (B4) it is evident
from their Eqs. (B7) and (B8) that `backreaction' (in the sense defined in the 
present paper) has to vanish identically: 
together with mass conservation (their Eq. (B10)) the introduction of
periodic boundary conditions already leads to the result that  
the scale factor $a_{\cal D}$ obeys the standard Friedmann equations.
(It should be noticed that ${\cal Q}_{\cal D} = 0$ is already {\it sufficient}
to have $a_{\cal D} (t) = a(t)$.)
Hence, according to {\it Corollary 3}, it is no surprise that the average 
Ricci scalar has to obey
the expansion law of models with spatial Ricci scalar $\propto a^{-2}$
in any consistent treatment of periodic perturbations on 
spatially flat space sections.  
Therefore, Eq. (3.1) of Russ et al. (loc.~cit.) cannot give any 
quantitative result
about the {\it global} value of the `backreaction' term: 
it vanishes by assumption and the
scale factor is given by the standard FRW cosmologies.
Note also that the introduction of a Fourier transform, or a decomposition into 
plane waves, respectively, is only meaningful in the case of spatially flat 
space sections, i.e., also the averaged Ricci scalar {\it has to} vanish.
It must be noted that the notion of `backreaction' as implied by 
Russ et al. (loc.~cit.) is slightly different from that in the present 
work. Any departure from the {\it flat} FRW cosmology in 
{\it curvature} (quantified by perturbation theory) may already be interpreted as
`backreaction' (Kasai, priv.~comm.).
\smallskip
This attempt illustrates the possibly cyclic nature of calculations of the 
`backreaction' term: if we start with spatially flat space sections
and a model for the inhomogeneous deformation tensor, the standard methods
of treating the perturbations as periodic on some scale already constrain the
cosmology to one without `backreaction'
(measuring the deviations from the family of FRW cosmologies). 
Note that in perturbation theory 
the first--order perturbations are sources of higher--order perturbations and,
as demonstrated by Russ et al. (loc.~cit.), a large class of 
periodic first--order and, in turn,
a large class of second--order perturbations
on a flat hypertorus average ${\cal Q}_{\cal D}$ to zero. 
\smallskip
In a realistic situation the domains on which one averages are not on average Ricci flat.
Large domains (e.g. of the size of the Universe) may not be easily compactified to make global statements about
the evolution of the Universe: non--trivial topological spaceforms 
have to be considered and simple periodic boundary conditions 
are no longer useful. For some further remarks see (Buchert 1997).

\bigskip\bigskip
\noindent
{\rma 3.5. Global structure versus local models}
\bigskip\noindent
The insight gained from the set of generalized Friedmann equations (10)
may be focussed in two ways: first, we are interested in the global 
structure of inhomogeneous cosmologies and, second, we would like to 
know more about average properties of spatial portions of the Universe
without severely restricting the dynamical model. 

\smallskip

As for the first point, 
a {\it globally} non--vanishing `backreaction' that may be small at early epochs
of the Universe' evolution could, on the scale of the size of the Universe,
slowly build up due to an amplification of inhomogeneities.
The more large--scale structure develops, the more the whole Universe 
might undergo global changes in morphology including the possibility of 
topology change. In order to analyze global changes  
during inflationary stages the present matter
model `dust' has to be generalized (which is the subject of a forthcoming
work, Paper II). At later epochs constraints from the microwave background
anisotropies can be used to give upper limits on the `backreaction' 
characteristic $\Omega_{\cal Q}$ (Eq. 12d): if we accept that the microwave
background dipol is only due to our proper motion against an isotropic
CMB reference frame, then limits on the global shear parameter
$$
\Sigma^2 : = {\langle\sigma^2 \rangle_{\rm CMB}\over 3 H_{\rm CMB}^2}
\eqno(15a)
$$
may be related by the assumption that, on the CMB scale,
$$
\theta \approx \langle\theta\rangle_{\rm CMB} \;\;;\eqno(15b)
$$
then, on the this scale,
$$
\Omega_{\cal Q} \approx \Sigma^2 \;\;.\eqno(15c)
$$
Maartens et al. (1995) have given upper limits on the shear parameter for
a Bianchi--type universe, in which case $\Sigma < 10^{-4}$ is a tight 
constraint on the global magnitude of the `backreaction' characteristic at the epoch
of last scattering
(see also: Wainwright \& Ellis 1997). 
As a disclaimer we note that an average over inhomogeneous models
is performed in Eq. (15a), and the average model must not 
necessarily isotropize as the Bianchi--type models (except type IX) do.
Therefore, care must
be taken in using such constraints at times after last scattering.

\smallskip

As for the second point, the expansion laws discussed in Appendix B and similar relations
(calculated, e.g., from perturbation theory, or hybrid models employing
perturbation theory on large scales, but including the full
nonlinearities on small scales, Takada \& Futamase 1999) 
provide a more general architecture
for the study of hierarchical cosmologies, understood in the sense of
models which do not single out
the non--generic case of scale--independent mean density, as the standard model
does. 
The focus here is on the {\it effective} dynamics of portions of the Universe
on some spatial scale including the possibility of statistically 
studying ensembles of spatial domains (see: Buchert et al. 1999 for an 
investigation within Newtonian cosmology).
With regard to the old ideas of hierarchical cosmologies 
the suggestion by Wertz (1971) may be put into perspective. It
relies on spherically symmetric domains which depend on their own 
parameters of a standard FRW cosmology. In this line
the expansion laws (B.3) (and more general relations) can also be applied to finite domains having their 
individual parameters. These parameters belong to generalized FRW cosmologies 
that include averages
over inhomogeneities encoded in an additional `backreaction' parameter 
(Eq. (12d)). The relevance of the characteristics (12) on a finite domain for
the interpretation of volume--limited observational data, where we cannot 
a priori assume that the surveyed volume is a portion of a standard Hubble
flow, is obvious. 

\smallskip

A similar view applies to so--called collapse models like the spherical ``top--hat''
model (e.g., Peebles 1980): on smaller spatial scales expansion fluctuations 
may become dominant in an overdense domain leading to collapse. The 
transition when ${\cal Q}_{\cal D}$ moves through zero as we come from large
scales can be used to mark the scale of
``decoupling of inhomogeneities from the global expansion''. 
In this context `backreaction' models
provide straightforward generalizations of the top--hat model. 
While Birkhoff's theorem lies at the basis of the spherical
model, the average dynamics including `backreaction' is not restricted by 
symmetry assumptions. Thus,
the averaged equations furnish a general framework with which 
one can describe the effective dynamics of 
individual collapsing or expanding domains.

\bigskip\bigskip
\smallskip
\noindent
{\rmb Acknowledgements:}

\noindent
I wish to 
thank J\"urgen Ehlers for his invitation to the `Albert--Einstein--Institut f\"ur Gravitationsphysik'
in Potsdam, where the body of this work was written during a visit in August 1997,
and to Gabriele Veneziano
for his invitation to CERN, Geneva, where it was completed. 
Special thanks to both and also 
to Mauro Carfora (Univ. of Pavia), Masumi Kasai (Univ. of Hirosaki), 
Martin Kerscher (Univ. of Munich) and Jean--Philippe Uzan (Univ. of Geneva)
for valuable discussions and comments.

\vfill\eject

\noindent{\rmm Appendix A: Newtonian Analogues}
\bigskip\bigskip
\noindent
The general expansion law in Newtonian cosmology reads (Buchert \& Ehlers 1997):
$$
3 {{\ddot a}_{\cal D} \over a_{\cal D}} + 4\pi G {M_{\cal D}\over V_{{\cal D}_o}a_{\cal D}^3 }
- \Lambda \;=\;{\cal Q}_{\cal D}\;\;\;.\eqno(A.1)
$$
As in the main text, $M_D$ denotes the total (conserved) rest mass contained in ${\cal D}$, 
and ${\cal Q}_{\cal D}$
is the same expression as Eq. (13c). This equation is identical to (10a),
as we already pointed out in (Buchert, Ehlers 1997).

The integral of Equation (A.1) with respect to time can be performed and 
yields the generalized form of 
Friedmann's differential equation for the first derivative of the
scale factor (Buchert 1996 -- with a different sign convention for 
${\cal Q}_{\cal D}$):
$$
{{\dot a}_{\cal D}^2 + k_{\cal D} \over a_{\cal D}^2 } - {8\pi G M_D \over 3 V_{{\cal D}_o}a_{\cal D}^3}
- {\Lambda \over 3}\;=\;{1\over 3 a_{\cal D}^2}\int_{t_0}^t \,dt' \;
{\cal Q}_{\cal D}\;
{d\over dt'}a^2_{\cal D}(t') \;\;\;,\eqno(A.2)
$$
where $k_{\cal D}$ is a (domain--dependent) integration constant.
\smallskip 
Comparing with the general relativistic equation (10b) we find 
the analogy:
$$
{k_{\cal D}\over a_{\cal D}^2 } - {1\over 3 a_{\cal D}^2}\int_{t_0}^t \,dt' \;
{\cal Q}_{\cal D}\; {d\over dt'}a^2_{\cal D}(t')
= {1\over 6}\left(\langle {\cal R} \rangle_{\cal D} + {\cal Q}_{\cal D}\right)
\;\;\;.\eqno(A.3) 
$$
The time--derivative of Eq. (A3) is equivalent to the integrability condition (13) in 
{\it Corollary 3}. We may view Eq. (A3) as an integral of (13b).
Eliminating the average curvature from this integral, Eq. (A3), and 
inserting it into the integrability condition (13b) formally results in a 
differential equation for ${\cal Q}_{\cal D}$ alone, which, however, reduces to an identity.
\smallskip
Notice that we cannot separately identify the integration constant 
$k_{\cal D}$ with the average Ricci scalar, since this would determine
the evolution of the average curvature and the `backreaction' term
to the special solution (B.3).
We might be able to show that, e.g., the solution (B.3) could also be found within the 
Newtonian framework for a special type of deformation of the domain's 
boundary. However, we cannot conclude that for the subclass of 
Newtonian cosmologies, which can be obtained from the Newtonian limit of the 
corresponding class of GR solutions, the solutions (B.3) would be 
the correct limit;
the limit ($c\rightarrow \infty$; ${\cal R} \rightarrow 0$) is not obvious
in the expression $c^2 {\cal R}$.

\vfill\eject

\noindent{\rmm Appendix B: Examples of Expansion Laws}
\bigskip\bigskip
\noindent
From Equation (13b) we conclude that, 

\noindent
first, for on average spatially flat domains the `backreaction' can be 
integrated to give
$$
{\cal Q}_{\cal D}(t) = {\cal Q}_{\cal D}^0 \;
a_{\cal D}^{-6} 
\;\;\;;\;\;\;{\cal Q}_{\cal D}^0 :={\cal Q}_{\cal D}(t_0)\;\;\;.\eqno(B.1a)
$$
Using the integral (B.1a) we write down a closed equation for the scale factor in 
this case:
$$
3{{\ddot a}_{\cal D} \over a_{\cal D}} + 4\pi G {M_{\cal D}\over V_{{\cal D}_o}a_{\cal D}^3} -
\Lambda \;=\; {{\cal Q}_{\cal D}^0 \over a_{\cal D}^6}
\;\;\;.\eqno(B.1b)
$$
In view of Eq. (10b) the integral of (B.1b) is given by 
$$
3\left( {{\dot a}_{\cal D}\over a_{\cal D}}\right)^2 - 8\pi G{M_{\cal D}\over V_{{\cal D}_o}a_{\cal D}^3}
 - \Lambda = -{{\cal Q}_{\cal D}^0 \over 2 a_{\cal D}^6} 
\;\;\;.\eqno(B.1c)
$$
Hence, the problem is reduced to a quadratur.

\bigskip\noindent
Second, for (on some spatial domain) vanishing `backreaction' we obtain
the special case of conformally constant curvature models;
the average scalar curvature is inversely proportional to the square of 
the ``radius of curvature'' on the domain,
$$
\langle {\cal R} \rangle_{\cal D} (t) =\langle {\cal R} \rangle_{\cal D}^0 \; 
a_{\cal D}^{-2} \;\;\;;\;\;\;\langle {\cal R} \rangle_{\cal D}^0 : = 
\langle {\cal R} \rangle_{\cal D} (t_0)\;\;\;,\eqno(B.2a)
$$
where $a_{\cal D}(t) = a(t)$ is a solution of the standard FRW models.
We are faced with the situation that the domain represents on average a small 
FRW universe with its own domain--dependent parameters.
However, here, this result holds for any spatial domain on which the 
`backreaction' vanishes, i.e. inhomogeneities are present and their 
fluctuations can even be large, but they have to compensate each other.

\bigskip\noindent
Third, one obvious solution of Eq. (13b) in the case of non--vanishing
average scalar curvature {\it and} non--vanishing `backreaction' is given by
$$
{\cal Q}_{\cal D}(t) = {\cal Q}_{\cal D}^0 \;
a_{\cal D}^{-6} 
\;\;\;;\;\;\;\langle {\cal R} \rangle_{\cal D} (t) =\langle {\cal R} 
\rangle_{\cal D}^0 \; a_{\cal D}^{-2} \;\;\;;\eqno(B.3a,b)
$$
here, $\langle {\cal R}\rangle_{\cal D}$ is proportional to an
``effective radius of curvature'' of the domain, 
and $a_{\cal D}(t) \ne a(t)$.
This solution features {\it independent} evolution of
average Ricci scalar and `backreaction': the spatial domain behaves
like a small ``almost FRW'' universe, still being characterized by its
own parameters, which are exclusively determined by the values of the 
fields {\it inside the domain}. Although special, this case offers 
the possibility of understanding some properties of the `backreaction' effect.
\bigskip
At first glance, it might look wrong that the dynamics of any patch of 
matter can be described independently of the environment; the non--local
gravitational influence from the matter outside the domain 
seems not to have impact on the dynamics of the domain. 
This interpretation is, however, misleading:
although we have to specify only initial data within the domain in, 
e.g., the solutions to (B.3), we still have to solve the constraint equations 
for these initial data which is a non--local operation and involves 
also the fields outside the domain under consideration. 
Still, solutions to (B.3) uniquely describe the averaged dynamics of the domain for 
all times, once the initial data are specified, without solving the constraints
at later times.

\bigskip

All of our examples  
restrict the generality: solution (B.1) is found by the
requirement of on the domain vanishing average Ricci scalar. Hence, in view
of the Hamiltonian constraint (4a) the inhomogeneities have to obey 
a special relation between the rest mass density, the cosmological constant 
and the second scalar invariant of the expansion tensor. 
Eqs. (B.3) together with the generalized Friedmann equations
determine a more general class of 
motions. Still, in general, we expect that the evolution of 
`backreaction' is coupled to the evolution of the averaged spatial curvature
in a more complex way.
Indeed, as Eq. (13a) shows, even for initially vanishing average 
curvature, there is generation of curvature in the 
course of structure formation, since amplification of
inhomogeneities builds up the `backreaction' term.

\bigskip

We finish the discussion of special expansion laws by 
giving a useful formula for the dynamical 
relation of the average characteristics (12). Combining (12) with 
{\it Corollary 3} and using the {\it Theorem} we obtain:
$$
{\dot\Omega}_{\cal Q} + 6H_{\cal D} \Omega_{\cal Q}
(1- \Omega_{k} - \Omega_{\cal Q}) + {\dot\Omega}_{k} + 
2H_{\cal D}\Omega_{k}(1- \Omega_{k} - \Omega_{\cal Q}) 
$$
$$
- 3H_{\cal D}
(1- \Omega_{\Lambda} - \Omega_{k} - \Omega_{\cal Q})
(\Omega_{k} + \Omega_{\cal Q}) \;=\;0\;\;.\eqno(B.4)
$$
One example may express a warning that the average characteristics in the present
case are dynamically tightly related and should not be treated independently:
let us assume that the cosmological term vanishes (which remains an
independent parameter), and that the restmass density 
characteristic remains constant in time mimicking the situation in an
Einstein--de Sitter universe. Then, in the simplest case ${\Omega_{m}} = 1$, 
the curvature
and `backreaction' characteristics have to compensate each other, 
$\Omega_{k}+\Omega_{\cal Q} = 0$, and Eq.~(B.4) reduces to 
$$
{\dot\Omega}_{\cal Q} + 6H_{\cal D}{\Omega}_{\cal Q} + {\dot\Omega}_{k} +
2H_{\cal D}\Omega_{k} = 0 \;\;.
$$
Eliminating one of the characteristics from this equation shows that 
the other has to vanish identically, reducing the average model
to the standard Einstein--de Sitter universe. 
There exists no generic inhomogeneous cosmology with $\Omega_m = 1$ throughout
its evolution. 

\vfill\eject

\noindent{\rmm Appendix C: Yodzis' Argument}
\bigskip\bigskip
\noindent
Let us view the averaging problem within a 4--dimensional tube of spacetime that is swept out
by a compact 3--dimensional domain $\cal D$ and bounded from
above and below by space--like hypersurfaces $t_1 = const.$, $t_2 =const.$ From the Ricci identity
$$
(u_{\alpha;\beta})_{;\gamma} - (u_{\alpha;\gamma})_{;\beta} =  
-R_{\alpha\delta\beta\gamma} u^{\delta}\;\;\;\eqno(C.1a)
$$
we obtain after contraction in $\alpha$ and $\beta$ and multiplication 
with $u^{\gamma}$:
$$
[u^{\alpha} u^{\beta}_{\;\,;\beta} - u^{\beta} u^{\alpha}_{\;\,;\beta}]_{;\alpha} = - R_{\alpha\beta}u^{\alpha}u^{\beta} +
(u^{\gamma}_{\;\,;\gamma})^2 - u^{\gamma}_{\;\,;
\delta}u^{\delta}_{\;\,;\gamma}\;\;\;.\eqno(C.1b)
$$
Recalling the definition of the second invariant of $K_{ij}$ (8b), 
we rewrite Eq. (C.1b):
$$
2 {\bf II} = [u^{\alpha}u^{\beta}_{\;\,;\beta} - u^{\beta}u^{\alpha}_{\;\,;\beta}]_{;\alpha} + 
R_{\alpha\beta}u^{\alpha}u^{\beta} \;\;\;.
\eqno(C.1c)
$$
After averaging we obtain for the `backreaction' term:
$$
{\cal Q}_{\cal D} = 2\langle {\bf II}\rangle_{\cal D} - {2\over 3} \langle {\bf I}\rangle^2_{\cal D} = 
$$
$$
{2\over V_{\cal D}}\int_{\cal D} J d^3 X \; [u^{\alpha}u^{\beta}_{\;\,;\beta} - u^{\beta}u^{\alpha}_{\;\,;\beta}]_{;\alpha}
-{2\over 3V_{\cal D}^2}\left(\int_{\cal D}J d^3 X \;u^{\beta}_{\;\,;\beta}\right)^2 + \langle R_{\alpha\beta}u^{\alpha}u^{\beta}
\rangle_{\cal D}
\;\;\;.\eqno(C.2)
$$
We now need to evaluate the total 4--divergences in Eq. (C.2) on the hypersurfaces. Yodzis (1974) gave the answer using 
the following argument: perform the volume integral over the 4--dimensional tube of spacetime $\Gamma$ and let then the 
distance between the hypersurfaces tend to zero, $t_2 - t_1 =\epsilon \rightarrow 0$. 
Applying Green's theorem on the integrals of the invariants,
$$
\int_{\Gamma(t_1, t_2)} d\Gamma\;u^{\alpha}_{\;\,;\alpha} =
$$
$$
 - \int_{{\cal D}_{t_2}}J d^3 X \;u^{\alpha}u_{\alpha} + 
 \int_{{\cal D}_{t_1}}J d^3 X \;u^{\alpha}u_{\alpha} = 
 V_{\cal D}(t_2) - V_{\cal D}(t_1)\;\;\;;\eqno(C.3a)
$$
$$
\int_{\Gamma(t_1, t_2)} d\Gamma\;[u^{\alpha}u^{\beta}_{\;\,;\beta} - 
u^{\beta}u^{\alpha}_{\;\,;\beta}]_{;\alpha} = 
$$
$$
- \int_{{\cal D}_{t_2}}J d^3 X \;[u^{\alpha}u^{\beta}_{\;\,;\beta} - u^{\beta}u^{\alpha}_{\;\,;\beta}] u_{\alpha} + 
 \int_{{\cal D}_{t_1}}J d^3 X \;[u^{\alpha}u^{\beta}_{\;\,;\beta} - 
 u^{\beta}u^{\alpha}_{\;\,;\beta}] u_{\alpha}\;\;\;,\eqno(C.3b) 
$$ 
he arrived at the following result by deviding Eqs. (C.3a,b) by $\epsilon$ and taking the limit $\epsilon \rightarrow 0$
(in our notations and conventions): first, he obtains the familiar equation 
(9d) for the average of the first invariant ${\bf I}=\theta$:
$$
\int_{\cal D} J d^3 X \; \theta = \langle\theta\rangle_{\cal D} 
V_{\cal D}\;\;\;;\eqno(C.3c)
$$
for the average of the second invariant ${\bf II}$ he derives: 
$$
\int_{\cal D} J d^3 X \; [- R_{\alpha\beta}u^{\alpha}u^{\beta} + 2 {\bf II}] = 
(\langle\theta\rangle_{\cal D} V_{\cal D})^{\bdot}\;\;\;.\eqno(C.3d)
$$
Although Yodzis (loc.~cit.) seems to imply that this result only holds for closed 3--spaces, we can immediately see
that these equations and especially Eq. (C.3d) hold for any compact and simply--connected domain in the hypersurfaces:  
from Einstein's equations (1a) we have $R_{\alpha\beta}u^{\alpha}u^{\beta} 
= 4\pi G\varrho - \Lambda$. Thus, Eq. (C.3d)
is equivalent to the averaged Raychaudhuri equation in {\it Corollary 1}, 
Eq. (11c). 
Calculating ${\cal Q}_{\cal D}$ from Eq. (C.2) we accordingly obtain Eq. (10a)
of the main text.

\bigskip\bigskip

\centerline{\rmm References}
\bigskip
\ref
Arnowitt R., Deser S., Misner C.W. (1962): in {\it Gravitation: an Introduction to Current Research}, L. Witten (ed.),
New York: Wiley
\ref
Buchert T. (1996): in ``Mapping, Measuring and Modelling the Universe'',
Val\`encia 1995, P. Coles, V.J. Mart\'\i nez, M.J. Pons--Border\'\i a (eds.), 
ASP conference series, 349--356. 
\ref
Buchert T. (1997): in ``2$^{nd}$ SFB workshop on
Astro--particle physics'', Report SFB/P002, Ringberg (Tegernsee) 1996, 
R. Bender, T. Buchert, P.
Schneider, F.v. Feilitzsch (eds.), 71-82.  
\ref
Buchert T., Ehlers J. (1997): {\it Astron. Astrophys.} {\bf 320}, 1.
\ref
Buchert T., Kerscher M., Sicka C. (1999): Preprint astro--ph/9912347.
\ref
Carfora M., Piotrkowska K. (1995): {\it Phys. Rev. D} {\bf 52}, 4393.
\ref
Ellis G.F.R. (1984): in {\it Gen. Rel. and Grav.}, B. Bertotti et al. (eds.),
Dordrecht, Reidel, 215--288.
\ref
Kasai M. (1995): {\it Phys. Rev. D} {\bf 52}, 5605.
\ref
Kofman L., Pogosyan D. (1995): {\it Ap.J.} {\bf 442}, 30.
\ref
Maartens R., Ellis G.F.R., Stoeger W.R. (1995): {\it Phys. Rev.} {\bf D51}, 
5942.
\ref
Matarrese S. (1996): in {\it Proc. IOP `Enrico Fermi'}, Course CXXXII 
(Dark Matter in the Universe), Varenna 1995, 
S. Bonometto, J. Primack, A. Provenzale (eds.), IOS Press Amsterdam, 601-628.
\ref
Matarrese S., Terranova D. (1996): {\it M.N.R.A.S.} {\bf 283}, 400.
\ref
Peebles P.J.E. (1980): {\it The Large Scale Structure of the Universe},
Princeton Univ. Press.
\ref
Russ H., Morita M., Kasai M., B\"orner G. (1996): {\it Phys. Rev. D} 
{\bf 53}, 6881.
\ref 
Russ H., Soffel M.H., Kasai M., B\"orner G. (1997): {\it Phys. Rev. D} {\bf 56}, 
2044.
\ref
Stoeger W.R., Helmi A., Torres D.F. (1999): Preprint gr--qc/9904020.
\ref
Takada M., Futamase T. (1999): {\it G.R.G.} {\bf 31}, 461.
\ref
Wainwright J., Ellis G.F.R. (1997): {\it Dynamical systems in cosmology},
Cambridge Univ. Press.
\ref
Wertz J.R. (1971): {\it Ap.J.} {\bf 164}, 227.
\ref 
Yodzis P. (1974): {\it Proc. Royal Irish Acad.} {\bf 74A}, 61.
\ref
York J.W. Jr. (1979): in ``Sources of Gravitational Radiation'', L. Smarr (ed.), 
Cambridge Univ. Press, p.83

\vfill\eject
\bye

\noindent{\rmm Appendix B: Solution of Equation (13d)}
\bigskip\bigskip
\noindent
As an illustration we consider the case $\Lambda = 0$.

\noindent
The quadratur (13d) is most conveniently solved for the variable
$V = |{\cal D}| = a_{\cal D}^3$; we first have
$$
dt = {dV \over\sqrt{C_1 V + C_2}} \;\;,\eqno(B.1)
$$
with $C_1 := 24\pi G M_{\cal D}$, 
$C_2 := -{3\over 2}{\cal Q}^0_{\cal D} (a^0_{\cal D})^6\;$.
We have to distinguish two cases:
\bigskip
\noindent\underbar{Case 1: $C_1 V + C_2 \ge 0$}
\bigskip\noindent 
Integrating (B.1) and solving for $V$ yields
$$
{V\over V_0} = 
\left[\alpha\left({t\over t_0}-1\right) + \sqrt{1 - \beta}\right]^2 + \beta
\;\;, \eqno(B.2a)
$$
with the dimensionless parameters ($V_0:=V(t_0)$)
$$
\alpha : = {t_0 \over 2}\sqrt{{C_1 \over V_0}}
= \sqrt{6\pi G \langle\varrho_0\rangle_{{\cal D}_0} t_0^2} 
\;\;,\;\;\beta : = -{C_2 \over C_1 V_0} = {{\cal Q}^0_{{\cal D}}\over 16\pi G 
\langle\varrho_0\rangle_{{\cal D}_0}}\;\;.
$$
In the case of vanishing `backreaction', $\beta = 0$, we have the solution
of the Einstein--de Sitter universe:
$$
{V\over V_0} = \alpha^2 \left[ {t\over t_0} - 1 \right]^2 + 
2\alpha \left[ {t\over t_0} - 1 \right] + 1 \;\;. \eqno(B.2b)
$$
Fixing the time--scale such that the model has a `Big--Bang' singularity,
i.e., $V(t \rightarrow 0) \rightarrow 0$,
gives $\alpha = 1$, and we obtain the model in its common normalization,
$$
{V\over V_0} = \left[ {t\over t_0} \right]^2 \;\;;\eqno(B.2c)
$$
hence, $a_{\cal D} \propto t^{2/3}$.

\smallskip\noindent
In the case of non--vanishing `backreaction' we normalize, for the sake of
comparison with the standard model, the time--scale to the same value
$\alpha = 1$. We then have for our solution:
$$
{V\over V_0} = 
\left[\left({t\over t_0}-1\right) + \sqrt{1 + \beta}\right]^2 - \beta
\;\;.\eqno(B.2d)
$$
Note that, with this normalization, the domain has a finite volume at $t=0$,
$$
V(t=0) = 2 V_0 (1-\sqrt{1 + \beta}) \;\;.
$$
It is interesting to see that global--in--time solutions admitting a 
`Big--Bang' singularity restrict possible values of the `backreaction' 
parameter to $\beta ...$. This can be understood as follows: negative values
of $\beta$ work in the sense of the mass density, while positive values
counteract the gravitational action of the mass: this is true for 
a critical value of beta....

The parameter choices $|\beta| = 1$ imply that the contribution of 
`backreaction' to the expansion is of the same order as the contribution
by the mass density, whereas $|\beta| = 0.25$ marks the equality of the 
respective contributions to the acceleration. 
\bigskip
\noindent\underbar{Case 2: $C_1 V + C_2 < 0$}
\bigskip\noindent 
There is a critical value of $\beta$ yielding recollapse
of the domain. In general the solution is oscillatory for ...

Consider the slope of ${\ddot a}_{\cal D}$ !

\vfill\eject
\bye